# Production of sub-gigabar pressures by a hyper-velocity impact in the collider using laser-induced cavity pressure acceleration

J. Badziak[1x], M. Kucharik[2], R. Liska[2]

[1]Institute of Plasma Physics and Laser Microfusion, 01-497 Warsaw, Poland

[2]Czech Technical University, FNSPE, 160 41 Prague 6, Czech Republic

Production of high dynamic pressure using a strong shock wave is a topic of high relevance for high energy density physics, inertial confinement fusion and materials science. Although the pressures in the multi-Mbar range can be produced by the shocks generated with a large variety of methods, the higher pressures, in the sub-Gbar or Gbar range, are achievable only with nuclear explosions or laser-driven shocks. However, the laser-to-shock energy conversion efficiency in the laser-based methods currently applied is low and, as a result, multi-kJ multi-beam lasers have to be used to produce such extremely high pressures. In this paper, the generation of high-pressure shocks in the newly proposed collider in which the projectile impacting a solid target is driven by the laser-induced cavity pressure acceleration (LICPA) mechanism is investigated using two-dimensional hydrodynamic simulations. A special attention is paid to the dependence of shock parameters and the laser-to-shock energy conversion efficiency on the impacted target material and the laser driver energy. It has been found that both in case of low-density and high-density solid targets the shock pressures in the sub-Gbar range can be produced in the LICPA-based collider with the laser energy of only a few hundreds of joules, and the laser-to-shock energy conversion efficiency can reach values of 10 – 20 %, by an order of magnitude higher than the conversion efficiencies achieved with other laser-based methods used so far.

## 1. Introduction

The generation of strong shock waves in condensed media has a long history and for decades high-pressure shocks have been used in the study of dynamic properties of matter under extreme conditions. Also presently, the production of high pressures using shocks is a hot topic developed in various branches of research, in particular in high energy-density physics (HEDP) (Drake, 2006; Nellis, 2006), inertial confinement fusion (ICF) (Atzeni and Meyer-ter-Vehn, 2004) or materials science (Nellis, 2006; Kraus et al., 2016). The shock-generated pressures in the multi-Mbar range can be achieved by a large variety of methods employing chemical (Fortov, 1982) or nuclear explosions (Trunin, 1994), heavy ion beams (Tahir et al., 2010), direct irradiation of a solid target by the laser beam (Veeser and Solem, 1978; Batani et al., 2002, 2014; Ozaki et al., 2004; Atzeni and Meyer-ter-Vehn, 2004; Guskov et al., 2007; Guskov, 2015) or the laser-produced X-rays (Lower et al., 1994; Benuzzi et al., 1996; Atzeni and Meyer-ter-Vehn, 2004; Hurricane et al., 2014) as well as the hyper-velocity impact, into a dense target, of a projectile accelerated by a pulsed-power machine (Lemke at al., 2005), a light-gas gun (Mitchell et al., 1991; Nellis, 2006) or a laser driver (Obenschain et al., 1983; Cauble et al., 1993; Ozaki et al., 2001; Karasik et al., 2010; Frantaduono et al., 2012; Shui Min et al., 2015; Min Shui et al., 2016). However, the pressures in the sub-Gbar or Gbar range are attainable currently only with laser-based methods (the nuclear explosions, which could also produce such pressures, are forbidden). Unfortunately, a common drawback of all laser-based methods of the shock generation used so far is the low efficiency of energy conversion from a laser to the shock, $\eta_{ls} = E_s/E_L$, which is below a few percent (Atzeni and Meyer-ter-Vehn, 2004; Cauble et al., 1993; Ozaki et al., 2001; Guskov et al., 2007; Guskov, 2015; Shui Min et al., 2015) ($E_s$ is the shock energy and $E_L$ is the energy of a laser beam which generates the shock or produces X-rays or accelerates the projectile). In case of the shock generation by the laser-driven hyper-velocity impact, the main reason for the low $\eta_{ls}$ value is the low energetic efficiency of projectile acceleration in the commonly used ablative acceleration (AA) scheme (employing either the laser or X-ray drive), which is $\eta_{acc} = E_p/E_L \sim 0.5 - 5\ \%$ (Cauble et al., 1993; Ozaki et al., 2001; Karasik et al., 2010; Frantaduono et al., 2012; Shui Min et al., 2015; Min Shui et al., 2016) ($E_p$ is the projectile kinetic energy). As a result, for the generation of sub-Gbar or Gbar shocks of spatial dimensions enabling their practical usefulness (e.g. studies of equations of state of various materials), multi-kJ laser drivers are necessary. The experiment carried out in the Lawrence Livermore National Laboratory, where for the generation of a quasi-planar, 0.4-mm, 750-Mbar shock in the gold target by the hyper-velocity impact, the impacting projectile (3-µm Au flyer foil) was accelerated in the AA scheme by soft X-rays produced by the ultraviolet beams of the NOVA laser of the total energy of 25 kJ (Cauble et al., 1993) could serve as an example. Higher shock pressures, in the multi-Gbar range, are achievable presently only in spherical shocks driven by many beams of large laser facilities (of energy from tens to hundreds kJ) used for ICF research (Atzeni and Meyer-ter Vehn, 2004; Hurricane et al., 2014).

---

[x] E-mail: jan.badziak@ifpilm.pl

Recently, a novel scheme of acceleration of dense plasma capable of accelerating a plasma projectile with the energetic efficiency of tens of percent has been proposed (Badziak et al., 2010, 2012). In this scheme - referred to as the laser-induced cavity pressure acceleration – LICPA - a projectile placed in a cavity is irradiated by a laser beam introduced into the cavity through a hole and then accelerated in a guiding channel by the pressure created in the cavity by the laser-produced hot plasma or by the photon pressure of the laser pulse trapped in the cavity. It has been demonstrated (Badziak et al., 2015, 2015a) that by the impact of the projectile driven by the LICPA mechanism into the solid target, an ultra-high-pressure shock can be generated in the target with the laser beam energy much lower than that required for producing such a shock with other laser-based methods used so far.

It should be emphasized that despite the apparent similarities of the LICPA scheme to the cannonball-like acceleration (CBA) scheme proposed in the early 80s (Azechi et al., 1981; Yamada et al., 1982; Fujita et al., 1985) (also known as the tamped ablation scheme (Ahlborn and Liese, 1981; Piriz and Tomasel, 1992)), these schemes are very different indeed. In the CBA scheme the plasma energy is efficiently accumulated in partially closed space – the cavity (like in the LICPA accelerator cavity) and then is gradually transferred to the target which closes the cavity. However, in planar geometry the energy transfer from the cavity to the target is effective only in the preliminary stage of the acceleration, that is until the target closes the cavity. When the target's movement causes its separation from the cavity (opening of the cavity), the energy stored in the cavity "escapes" in the free space (like the energy stored in the balloon after pricking it with a pin). Moreover, since the target (projectile) moves in the free space, the internal energy accumulated in it (generated as a result of its heating with a shock wave) leads to its spatial expansion both in the longitudinal and radial direction. Consequently, there is a large increase in the volume of the projectile and significant decrease of its density. What even seems possible is the burst of the projectile into tiny pieces (in case when the energy transferred from the cavity to the target is of significant amount). Due to the low density of such a projectile, its large final thickness and relatively low speed (limited by the short time of efficient acceleration) and its possible considerable inhomogeneity its application, e.g. to generate shock waves by the projectile impact into a solid target is inefficient. Probably, due to the above-mentioned disadvantages, the CBA scheme has not found a wider application neither in the inertial fusion nor in other branches of HEDP despite the fact that the energetic efficiency of projectile acceleration in that scheme can be higher than in the AA scheme (Azechi et al., 1981; Yamada et al., 1982; Fujita et al., 1985). In the LICPA scheme, the situation is fundamentally different from the one in the CBA scheme. Thanks to the existence of the guiding channel in the LICPA accelerator, firstly, the process of acceleration can last much longer than in case of CBA and almost all energy accumulated in the cavity can be transferred to the projectile (accelerated target); secondly, the radial expansion of the projectile does not take place, and the longitudinal expansion is significantly reduced. As a result, the projectile in the output of the LICPA accelerator channel has higher velocity, considerably smaller thickness and much higher density and kinetic energy fluence than in the CBA scheme. Consequently, e.g. the shock wave generation by the projectile impact into a solid target is much more efficient than in case of CBA. The above-indicated differences between CBA and LICPA were proved in our experiment at the kilojoule PALS laser facility where we measured craters produced in the Al massive target by the impact of the Al projectile accelerated by LICPA or by CBA as well as plasma fluxes leaving the LICPA and CBA accelerators using optical interferometry.

In this paper, the process of shock generation in a solid target by the impact of projectile accelerated in the LICPA scheme is investigated in detail with the use of numerical hydrodynamic simulations. A particular attention is paid to the dependence of shock parameters and the laser-to-shock energy conversion efficiency on the impacted target material and the laser driver energy. It is shown that both in case of low-density (Al, polystyrene-CH) and high-density (Cu, Au) solid targets the shock pressure in the sub-Gbar range can be generated in the LICPA-based collider with the laser driver energy of only a few hundreds of joules and the laser-to-shock energy conversion efficiency can reach values above 10 %.

## 2. The LICPA-based collider for shock generation

The idea of a LICPA-driven collider in cylindrical (a) and conical (b) geometry is presented in Figure 1. In the collider, a projectile (e.g. a heavy disc covered by a low-Z ablator) placed in a cavity is irradiated by a laser beam introduced into the cavity through a hole and then accelerated in a guiding channel by the pressure created in the cavity by the laser-produced hot plasma expanding from the irradiated side of the projectile or by the photon (radiation) pressure of the ultra-intense laser pulse trapped in the cavity. The projectile accelerated in the channel impacts into a dense target placed at the channel exit. Due to the impact, the projectile energy and momentum are rapidly transferred to the target and when the projectile velocity significantly surpasses the sound velocity in the target, a high-pressure shock is generated in the target.

As compared to other laser-based schemes of shock generation, in particular those using the hyper-velocity impact for the shock generation, the LICPA-driven collider has several significant advantages, in particular:

a) very high energetic efficiency of projectile acceleration, $\eta_{acc} = E_p/E_L$, which can reach tens of percent both for a short-wavelength (< 0.5μm) and a long-wavelength (~1μm) laser (Badziak et al., 2012);

b) high density of the projectile at the moment of the impact, which ensures the efficient transfer of the projectile energy to the impacted (dense) target and the generated shock (Badziak et al., 2012, 2015a);

c) capability of accelerating the projectile to very high velocities; in the hydrodynamic acceleration regime (driven by the hot plasma pressure) (Badziak et al., 2012), the projectile velocity can be a few times higher than in the conventional ablative acceleration (AA) scheme (the temperature of plasma in the cavity can be up to a factor ~ 10 higher than that of the freely expanding plasma) and can potentially reach the value well above 1000 km/s, while in the photon-pressure acceleration regime, the projectile can be accelerated even to sub-relativistic velocities (Badziak et al., 2012);

d) existing of additional propelling of the projectile during and after the collision by the pressure still stored in the guiding channel;

e) no significant pre-heating of the impacted target by X-rays or hot electrons (especially when they are produced from a low-Z ablator), since the high-density projectile can effectively absorb them.

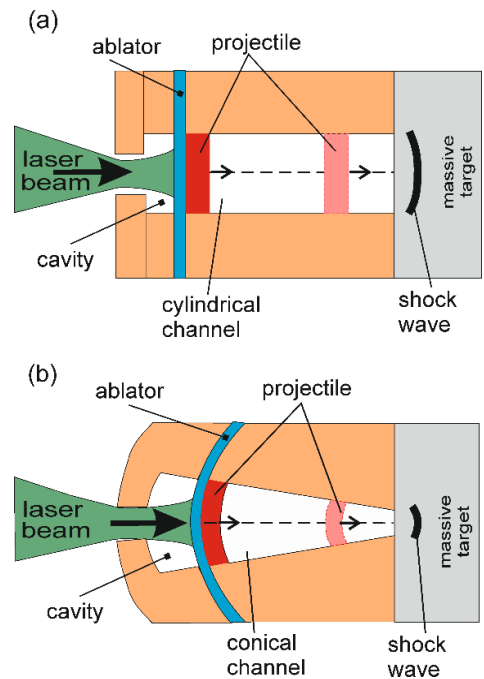

Fig. 1. Simplified scheme of the LICPA-based collider in cylindrical (a) and conical (b) geometry. The shock wave is generated in the massive target due to the impact of the projectile accelerated in the collider guiding channel into the target.

Taking into account the above-mentioned properties of the LICPA-based collider, it could be expected that the laser-to-shock energy conversion efficiency $\eta_{ls}$ in the collider will be even by an order of magnitude higher than for other schemes of shock generation used so far and, as a result, the laser energies needed for generating the shocks of the required energy or pressure will be much lower than the energies needed in other schemes.

In this paper, we investigate the properties of shocks produced in the cylindrical LICPA-based collider (Fig. 1a) working in the hydrodynamic regime. The shocks are generated in the solid massive targets made of various materials by the impact of a gold plasma projectile driven by the pressure of plasma produced in the collider cavity by a high-intensity (~ $10^{15}$ W/cm$^2$) sub-nanosecond laser pulse of energy below 0.5 kJ.

## 3. Results of numerical simulations and discussion

For numerical investigations of shock generation by a hyper-velocity impact in the LICPA-based collider we used the two-dimensional (2D) hydrodynamic PALE (Prague Arbitrary Lagrangian Eurelian) code (Kapin et al., 2008; Liska et al., 2011). The computations encompassed both the projectile acceleration phase and the phase of the projectile impact into the solid massive target placed at the LICPA accelerator channel exit. In particular, they included: the interaction of the laser beam with the projectile ablator and the absorption of the laser radiation in the plasma produced and confined in the cavity, the shock generation in the projectile (by the plasma pressure) and its heating and acceleration in the guiding channel, the impact of the projectile into the massive target and the generation of the shock in the target and, finally, heating the target by the shock, melting and evaporating of the target material and, as a result, the formation of a crater in the target. The simulations of the acceleration phase were carried out in Lagrangian coordinates applying a mesh moving with the plasma. In turn, in the impact phase, the simulations were performed in the Eulerian coordinates with a static computational mesh as the computational domain in this phase does not change. In this phase, the computations had been conducted for a long time (typically, for ~ 1000 ns) up to the moment when parameters (pressure, temperature, velocity) of the shock, generated shortly after the impact and propagating in the target, became too low to melt the target material, and the dimensions of the crater produced as a result of vaporization and melting of the material stopped to increase. Both in the acceleration phase and the impact phase, the QEOS equations of state (More et al., 1988) were applied to describe thermodynamic properties of materials used in the projectile and the impacted targets. For both phases computations were performed in the cylindrical r, z geometry (for more details about the code and the simulations see (Kapin et al., 2008; Liska et al.,2011)).

The simulations were performed for the same LICPA accelerator parameters as the ones used in the experiment carried out recently at the PALS laser facility (Badziak et al., 2015a). The accelerator walls were made of gold and the main accelerator dimensions were as follows: the cavity length $L_c$ = 0.4 mm, the channel length $L_{Ch}$ = 2 mm, the cavity diameter $D_c$ = 0.3 mm, and the channel diameter $D_{Ch}$ = 0.3 mm. As a projectile in its initial state we applied the gold disc of the thickness of 2.8 um and the diameter of 0.3 mm (the projectile mass $m_p$ ≈ 4 μg), covered by the CH ablator of the thickness of 5 μm. The massive targets placed at the accelerator channel exit were made of Au, Cu, Al or CH. The laser beam parameters on the target (CH ablator) were assumed to be relevant to the parameters of the 3ω PALS laser beam, in particular: the laser wavelength λ = 0.438 um, the pulse duration $\tau_L$ = 0.3 ns, the transverse beam intensity distribution ~ $\exp(r/r_o)^6$ with $r_o$ = 170 μm. For the laser energy $E_L$ = 200 J, these parameters correspond to the beam peak intensity on the target ~ $10^{15}$ W/cm$^2$. The laser beam energy varied within the range 10 – 400 J.

The correctness and accuracy of the PALE code used for the simulations was verified for various laser-based schemes of shock generation, in particular for the LICPA-based collider. In that case, the verification was made by the comparison of parameters of craters produced in a massive Al target by the impact of Al projectiles of various masses calculated using the code with the ones obtained from measurements performed at the PALS laser facility, and a very good qualitative and quantitative agreement between the numerical and experimental results was found (Badziak et al., 2015a). The code was also used to simulate the shock generation in a CH/Cu target by the direct target irradiation by the PALS laser beam and fairly good agreement (within ~ 30 %) between the results of the PALE simulations and the results of measurements was obtained as well (Koester et al., 2013; Badziak et al., 2015b). Positive results of these verifications allow us to believe that numerical results obtained in this paper are correct not only qualitatively but also quantitatively within a factor ~ 1.5 or smaller.

Fig. 2 presents 2 D spatial profiles of density (a), temperature (b), pressure (c) and axial (along z) velocity (d) of plasma inside the channel of the cylindrical LICPA-based collider at a final stage of acceleration of the gold projectile, i.e. at the moment when the projectile starts to leave the channel and begins to interact with the massive target placed at the channel exit (at z = - 2000 μm).

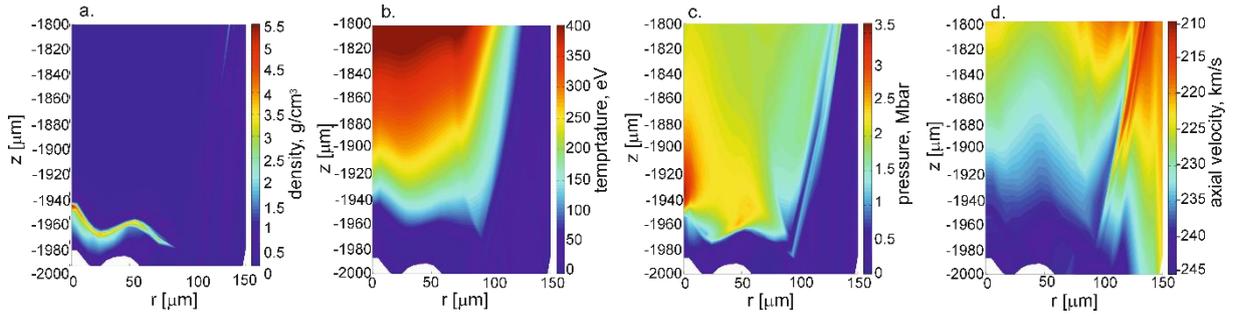

Fig. 2. 2D spatial profiles of density (a), temperature (b), pressure (c) and axial velocity (d) of plasma inside the end portion of the collider channel at the moment when the gold projectile starts to leave the channel and begins to interact with the massive target situated at the channel exit (z = - 2000 um). The CH/Au target placed at the beginning of the channel (z = 0) is irradiated by the laser beam introduced into the collider cavity and directed along the channel axis (r = 0). $E_L$ = 200 J.

It can be seen that at this stage the projectile is a fairly compact and dense plasma object of the effective thickness below 10 μm and the density averaged within this thickness ~ 2 – 3 g/cm$^3$. The projectile moves with the mean velocity above 200 km/s and the majority of the projectile mass and kinetic energy is accumulated in the region of Δz < 20 μm, Δr < 100 μm. The projectile surface is not ideally flat and homogeneous, so we can expect that also the shock produced in the massive target by the projectile impact will not be ideally homogeneous in space. It is also worth noticing that there is still a large amount of energy stored in the channel behind the projectile (mostly the energy of the ablating CH plasma) which has not been transformed into the projectile yet (Fig. 2c). In case of using a longer guiding channel (which ensures a longer acceleration time) this energy could be converted into the kinetic energy of the projectile increasing its total energy and, as a result, increasing the energy and pressure of the generated shock. However, even if the energy stored in the channel is not transformed completely into the projectile energy during the projectile acceleration, a part of the remaining energy can still be converted into the shock energy after the projectile impact. How much of the ablating plasma energy is transformed into the projectile kinetic energy depends on the time of projectile acceleration and can be to some extend controlled by changing the channel's length. More detailed discussion of the process of projectile acceleration in the LICPA scheme can be found in our previous papers (Badziak et al., 2012, 2015, 2015a).

Further, we will show how parameters and the structure of the shock generated in a massive target by the projectile impact is influenced by the properties of the target material and especially by the material density. Figs. 3, 4 and 5 present 2D spatial distributions of the density, the pressure and the temperature in the impacted target (z < 0) made of Au, Cu, Al or CH at the moment t = $t_{pmax}$ when the pressure in the target attains the highest value. Although maximum values of these parameters in the generated shock significantly depend on the kind of the material, for all target materials the spatial shape of the shock is similar – the shock is rather flat and homogeneous within 2r = 200 μm (in spite of the fact that the projectile surface is pleated). The quasi-planar shape of the shock is preserved up to ~ 1 ns after reaching by the shock the highest pressure, and after this period a radial expansion of the shock begins to be significant. The last effect can be seen in Figs. 6 and 7 where

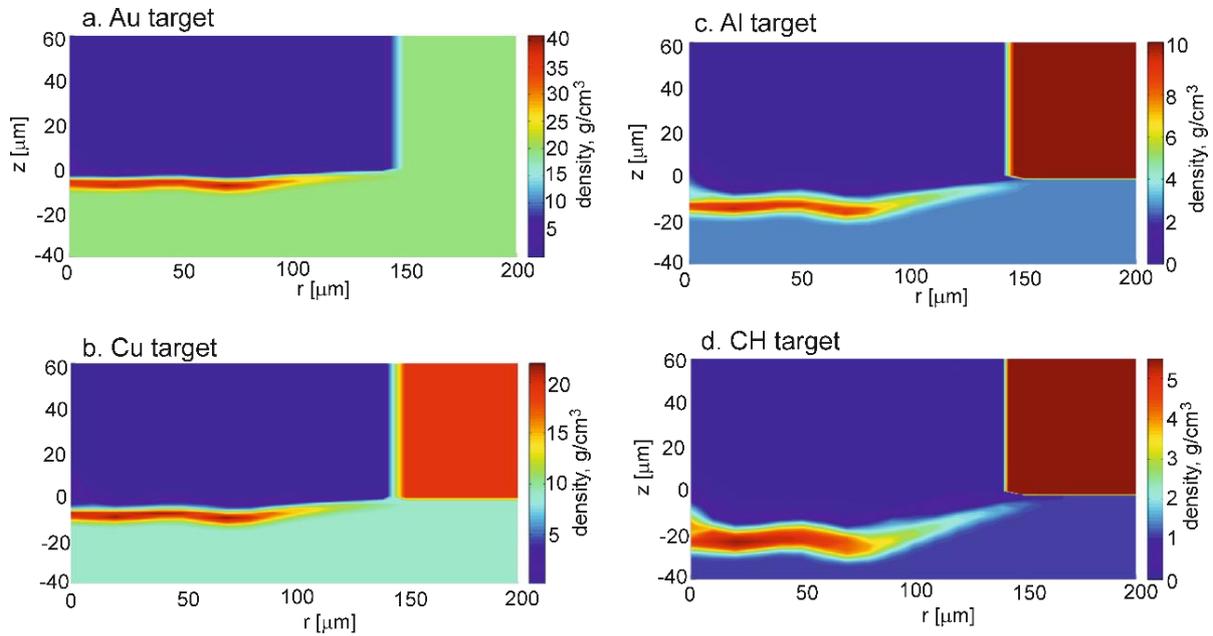

Fig. 3. 2D spatial profiles of density inside targets (z ≤ 0) made of various materials at the moment when the pressure in the target attains the highest value. The targets are impacted by the gold projectile accelerated in the collider channel situated in the region z > 0, r < 150 um. $E_L$ = 200 J.

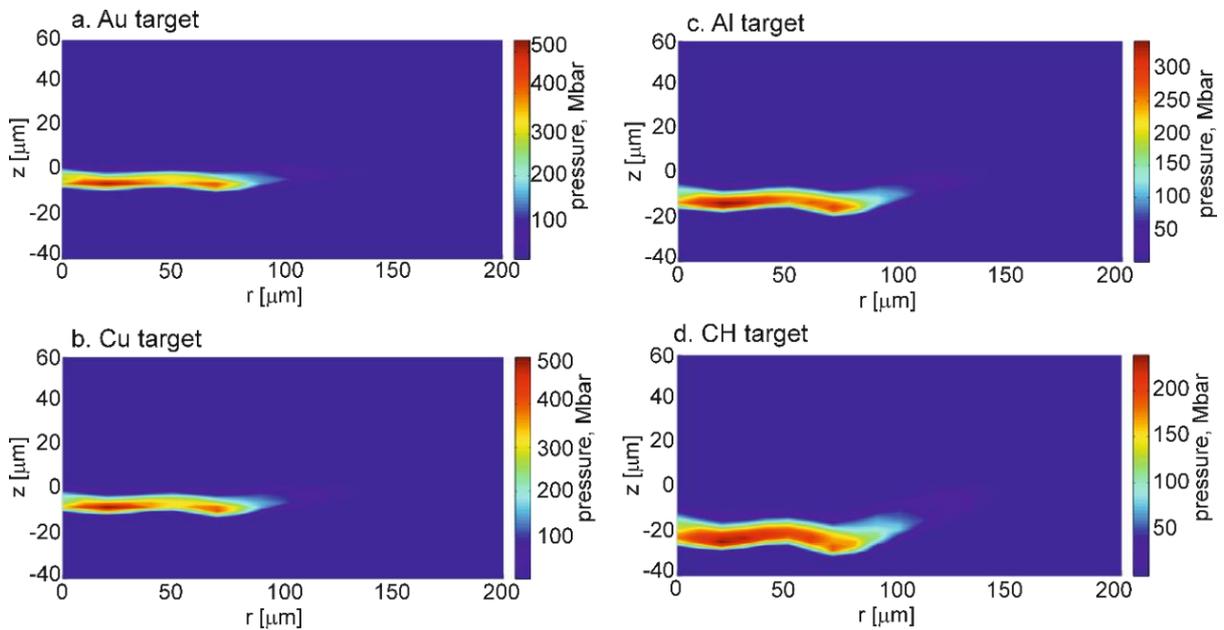

Fig. 4. 2D spatial profiles of pressure inside targets (z ≤ 0) made of various materials at the moment when the pressure in the target attains the highest value. The targets are impacted by the gold projectile accelerated in the collider channel situated in the region z > 0, r < 150 um. $E_L$ = 200 J.

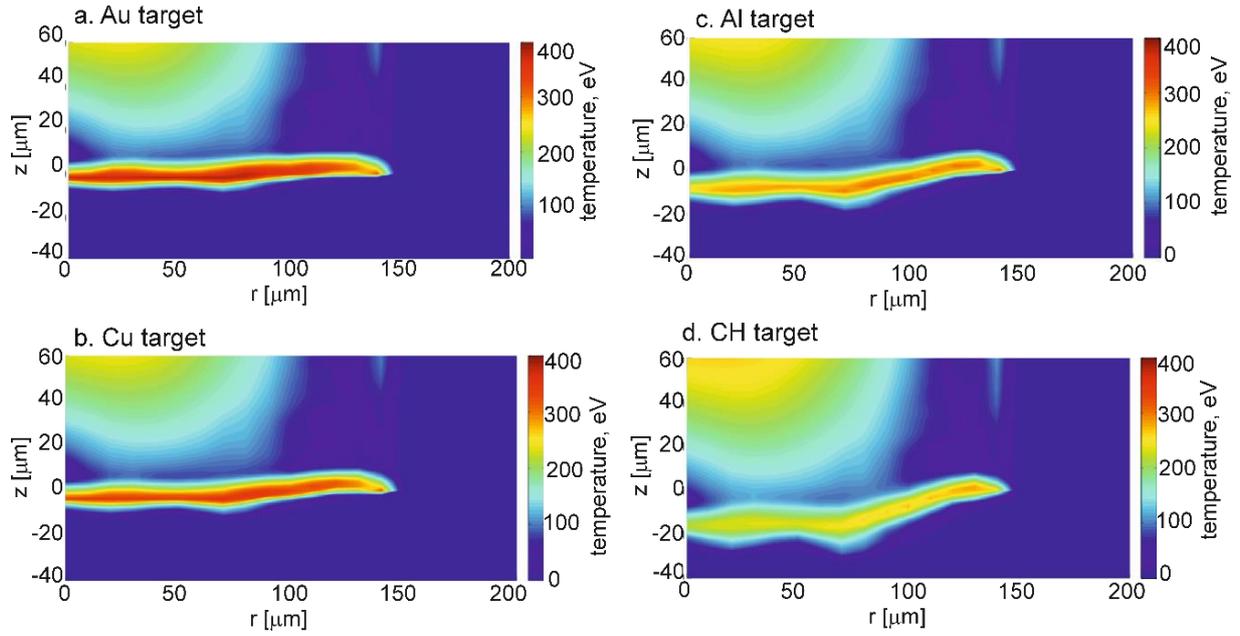

Fig. 5. 2D spatial profiles of temperature inside targets (z ≤ 0) made of various materials at the moment when the pressure in the target attains the highest value. The targets are impacted by the gold projectile accelerated in the collider channel situated in the region z > 0, r < 150 um. $E_L$ = 200 J.

2D spatial distributions of the density and the pressure in the targets 5ns after the projectile impact is presented. In this moment, the shock front is not flat and the pressure and density in the shock is distributed less homogenously than at t = $t_{pmax}$. Since during propagation of the shock in the target the volume of the shock increases and the energy of the shock is transferred to the target material, the density in the shock and the shock pressure decrease. The material heated by the shock is melted and evaporated and, as a result, a crater is formed in the target. The crater volume is proportional to the shock energy (Guskov et al., 2007) and its shape reproduces the shape of the shock which in the late propagation phase takes a quasi-spherical shape. It is demonstrated in Fig. 8, where 2D spatial distributions of the temperature in the impacted targets at the moment t = 1000 ns after the impact is shown. The bright quasi-spherical line in the figure represents the shape of the shock and, simultaneously, a border of the crater produced in the target.

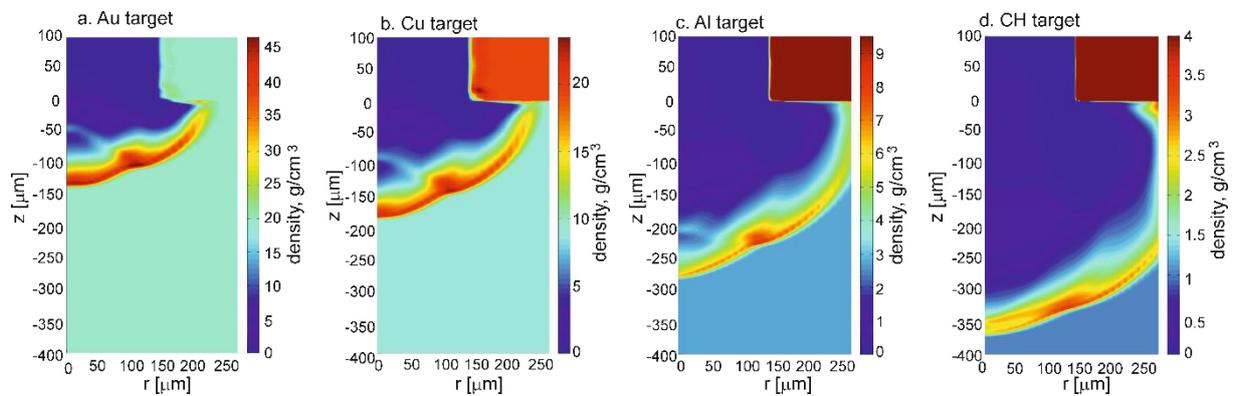

Fig. 6. As in Fig. 3, but at the moment of 5 ns after the projectile impact.

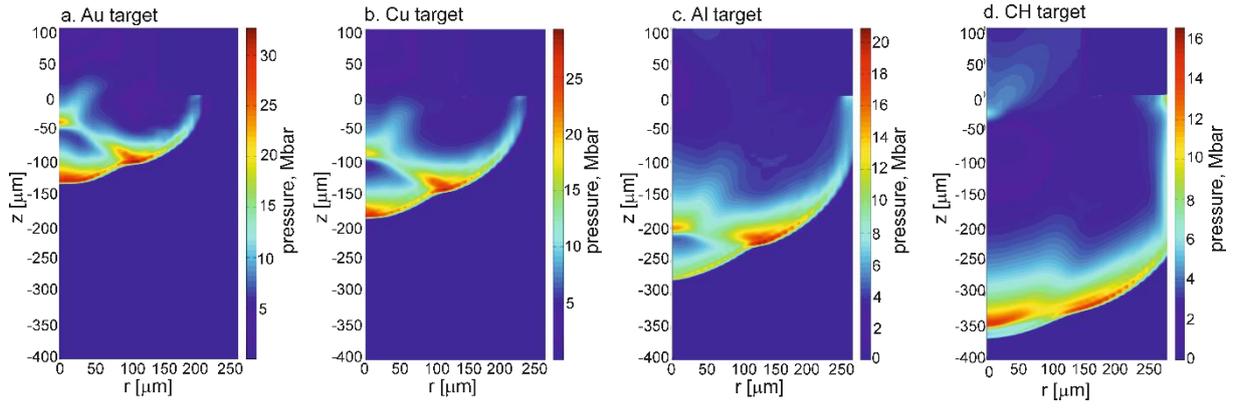

Fig. 7. As in Fig. 4, but at the moment of 5ns after the projectile impact.

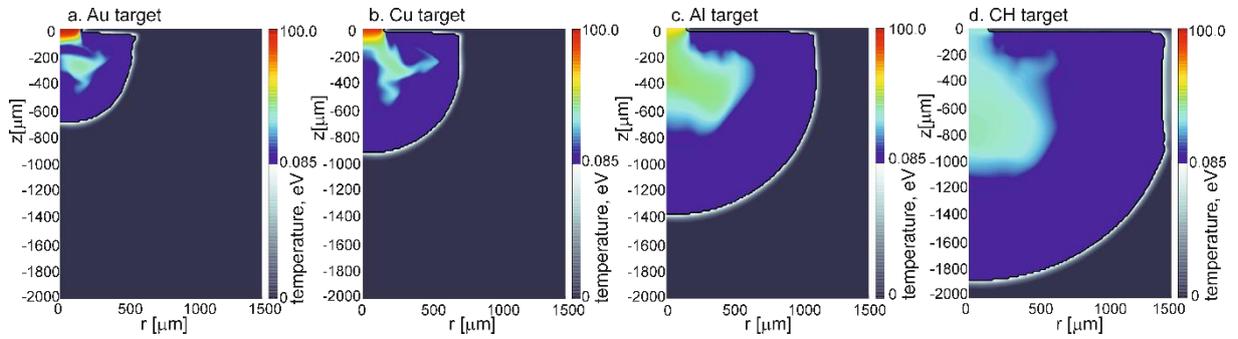

Fig. 8. 2D spatial distributions of temperature inside targets (z ≤ 0) made of various materials at the final stage of formation of crater in the target (t = 1000 ns). The targets are impacted by the gold projectile accelerated in the collider channel situated in the region z > 0, r < 150 um. $E_L$ = 200 J.

The highest value of pressure in the impacted target changes in time. Fig. 9 shows how the maximum pressure in the Au, Cu, Al or CH target changes within the time period of 5 ns after the projectile impact. As can be seen, the maximum pressure (in the target space) attains its peak value very fast (within ~ 0.2 ns) independently of the target material. It indicates a very short time of energy transfer from the projectile to the target, which is a consequence of the high velocity and compactness (small effective thickness) of the projectile. The highest peak pressure is achieved for the target of the highest density (Au target); however, a difference between the highest pressures in Au and Cu is rather small despite a factor 2 higher density of Au. This is due to the fact that the shock pressure depends not only on the impacted target density but also on other target material properties, in particular the pressure decreases with an increase in the material compressibility, which is higher for Au than for Cu. In the time period of Δt = 1 ns after the impact, when planarity of the shock is preserved, the maximum shock pressure averaged in time is by a factor 1.5 – 2 lower than the peak pressure and lays in the sub-Gbar range both for the high-density and low-density targets used in the simulations. This period is by a factor ~ 2 longer than that estimated for the sub-Gbar pressure range in the hyper-velocity impact experiment with the multi-kJ Nova laser [19] and seems to be long enough to measure some parameters of the material compressed by the shock, e.g. those required for the determination of equation of state of the material.

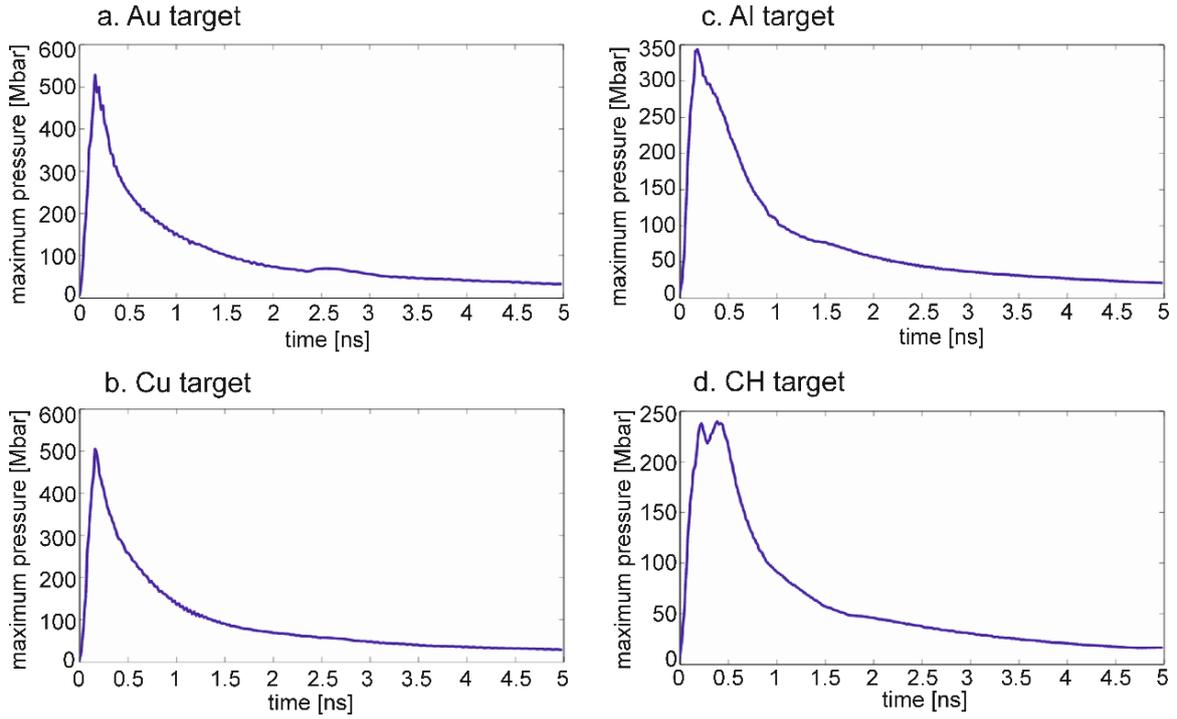

Fig. 9. The temporal dependence of the maximum pressure inside targets made of various materials impacted by the gold projectile. $E_L$ = 200 J.

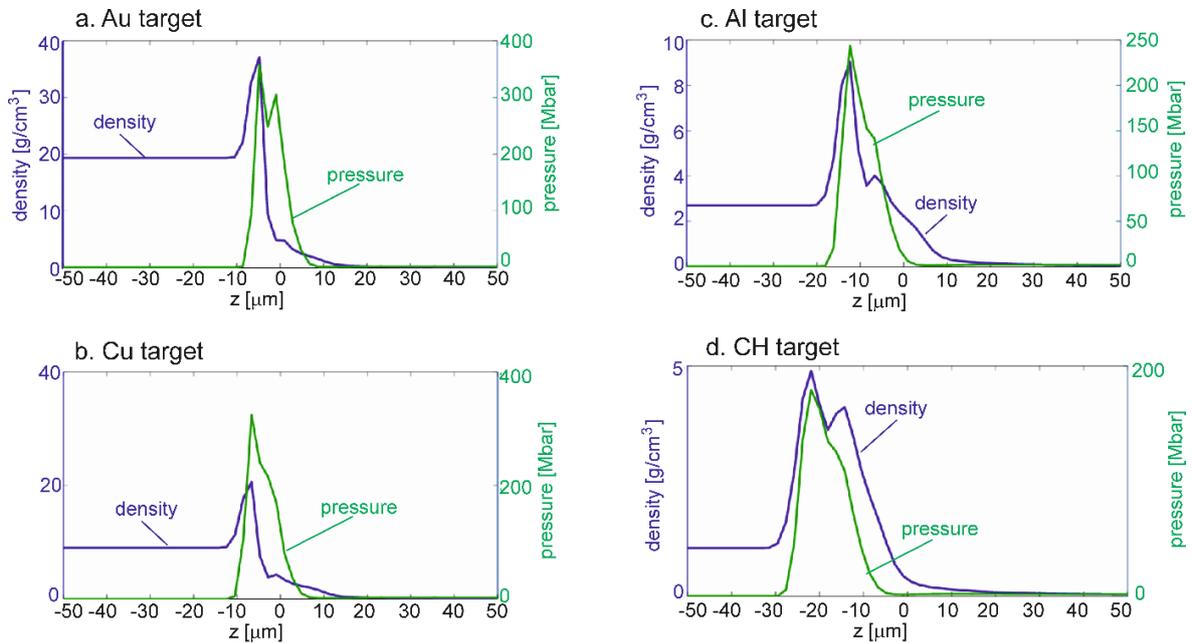

Fig. 10. Spatial profiles of density and pressure in the impacted Au, Cu, Al or CH target along the shock propagation axis (at r = 0) for the time t corresponding to the peak pressure. $E_L$ = 200 J.

Fig. 10 presents spatial profiles of density and pressure in the impacted Au, Cu, Al or CH target along the shock propagation axis (z) at r = 0 for the time t corresponding to the peak pressure. For all the targets, the peak pressure is reached at the point z where also the density attains the peak value;

however, the effective thickness of the shock depends on the target density and is the largest for the CH target. The target material in front of the shock is not disturbed and no preheating of the target due to the projectile impact is observed. It can be also expected that this material is not disturbed/preheated by X-rays and hot electrons produced at the laser-plasma interaction inside the collider cavity, since the high-density (of a high density-thickness product $\rho l$) projectile effectively absorbs both the X-rays and the hot electrons. In addition, low-Z ablator helps to avoid the preheating by decreasing the efficiency of the X-rays and hot electrons production and lowering X-ray and electron energy. This is an additional advantage of the LICPA-driven collider relative to the scheme with a direct laser irradiation of a solid target and also in comparison with the case when the target is impacted by a projectile driven by ablative acceleration in free space. In the last case – due to the longitudinal and transverse expansion of the projectile plasma during acceleration - the value of $\rho l$ for the impacting projectile is much lower than in the case of projectile driven by LICPA and, as a result, the laser-produced X-rays and hot electrons are less effectively suppressed by the projectile.

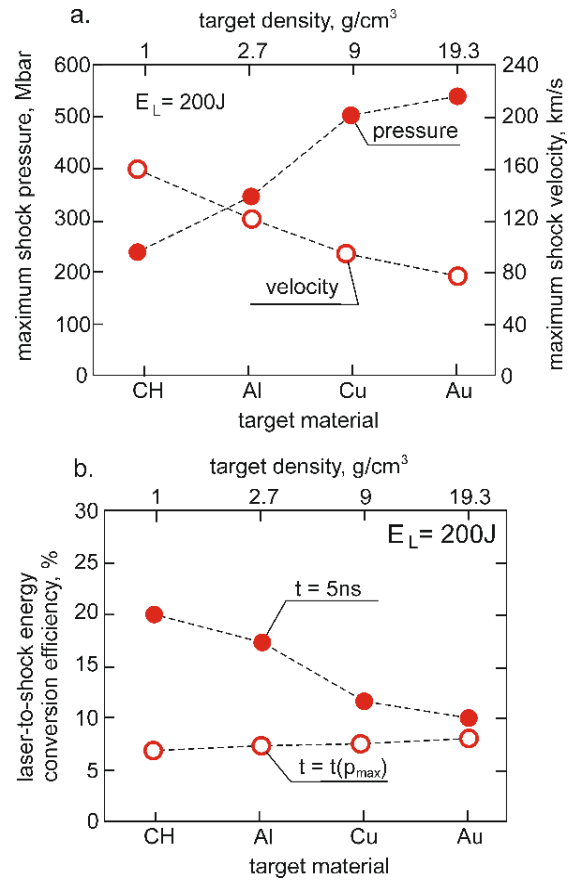

Fig. 11. The maximum values of pressure and velocity of the shock generated inside the targets made of various materials impacted by the gold projectile (a). The laser-to-shock energy conversion efficiency for the targets made of various materials at the moment $t = t_{pmax}$ when the shock pressure achieves the highest value and at the moment of 5 ns after the gold projectile impact (b). $E_L$ = 200 J.

The summary quantitative results of the simulation of shock generation by the projectile impact in the targets made of different materials are shown in Fig. 11. Fig. 11a presents the maximum (in space and time) pressures and velocities of the shocks attained in the targets and Fig. 11b the laser-to-shock energy conversion efficiency $\eta_{ls} = E_s/E_L$ for the time $t = t_{pmax}$ when the pressure reaches the highest value and for t equal to 5ns after the impact. The shock energy $E_s$ is defined here as the energy of the target material which is flowing along the axis z into the target interior with the velocity higher than ¼ of the maximum flow velocity. The maximum shock velocity decreases with an increase in the target material density while the maximum shock pressure increases with the material density and it approaches 500 Mbar for Au and Cu and surpasses 300 Mbar for Al. So far, such high values of the pressure could be produced with the laser-based methods only with the use of large multi-kJ lasers. Here, the sub-Gbar shock pressures are achieved at laser energy of only 200 J. Attaining such high pressures in the LICPA-based collider at relatively small laser energy is possible thanks to the very high efficiency of the projectile acceleration by the LICPA mechanism, high efficiency of the energy transfer from the projectile to the target due to high projectile density and a short interaction time between the projectile and the target (being the result of compactness and high velocity of the projectile). The first two factors also determine the high value of the laser-to shock energy conversion efficiency which for low-density targets (CH, Al) approaches 20 % and is by an order of magnitude higher than the efficiencies attained with laser-based methods so far.

The results presented in Fig. 11 are in a qualitative agreement with a simple steady-state model of the shock generation in a target of density $\rho_t$ impacted by a projectile of density $\rho_p$ moving with velocity $v_p$ (e.g. Gehring, 1970]). According to this model, the pressure of the shock generated in the target is determined by the expression (Gehring, 1970):

$$p = \left(\frac{\gamma_p+1}{2}\right) * \frac{\rho_p v_p^2}{(1+\beta^{1/2})^2} \, , \qquad (1)$$

where $\beta = (\gamma_p +1) \rho_p /(\gamma_t +1) \rho_t$ and $\gamma_t$, $\gamma_p$ are the adiabatic exponents of the target and the projectile materials, respectively. It can be shown from (1) that at a fixed projectile density, an increase in the target density results in an increase in the shock pressure, in accordance with what we observe in Fig. 11a. On the other hand, when the target density is fixed, the shock pressure increases with an increase in the projectile density. Based on this model, the ratio $\alpha$ of the energy transferred to the target by the projectile impact and the projectile kinetic energy can be expressed by (Gehring, 1970; Guskov et al. 2009):

$$\alpha = \frac{2\beta^{1/2}}{1+\beta} \, , \qquad (2)$$

The ratio $\alpha$ reaches a maximum value when $\beta = 1$, so when the projectile density is close to the target density (at the assumption that $\gamma_t$ and $\gamma_p$ do not differ considerably). In such a case, half of the projectile energy is transferred to the shock generated in the target. The expression (2) qualitatively explains the dependence of the laser-to-shock energy conversion efficiency on the target density observed for t = 5ns in Fig. 11b. Since the effective density of the projectile impacting the target $\rho_p$ ~ 2 – 3 g/cm$^3$ is rather close to the initial density of CH and Al, the efficiency of the energy transfer from the projectile to these targets is close to the maximal one and higher than for Au and Cu which densities are considerably higher than the projectile density. As a result, also values $\eta_s$ for CH and Al are higher than the ones for Au and Cu. Although the expressions (1) and (2) explain qualitatively dependencies of the shock parameters on the impacted target density, the absolute values of these parameters calculated from these expressions are far from those obtained from the simulations. This is due to the fact that the process of shock generation in the LICPA-based collider is much more complex than that assumed in this very simple model. In particular, the model does not take into account the fact that a significant influence on the shock parameters, especially in the later phases of shock generation (t > 1ns), has the energy of plasma stored in the collider channel behind the projectile which is still transferred to the shock after the impact. This energy can be comparable to the projectile energy and, as a result, its transfer to the shock can be a factor which remarkably increases the shock energy.

The maximum (in space and time) pressure of the shock and the laser-to-shock energy conversion efficiency, $\eta_{ls}$, as a function of the laser energy, for the Al massive target impacted by the gold projectile, are presented in Fig. 12. The shock pressure increases almost linearly with the laser energy at the energy range of 10 – 400 J accessible at the PALS facility and at $E_L$ = 400 J it attains 0.7 Gbar. The conversion efficiency decreases with an increase in the laser energy in the low energy range (below of 100 J) and then slightly increases. At $E_L$ = 400 J, the conversion efficiency reaches 10% at the early phase of the shock formation (when the shock pressure reaches the maximum value) and approaches 20% in the late phase (at t = 5ns) of the shock propagation (not shown in the figure). The demonstrated conversion efficiencies are by an order of magnitude higher than those achieved with other laser-based methods used so far.

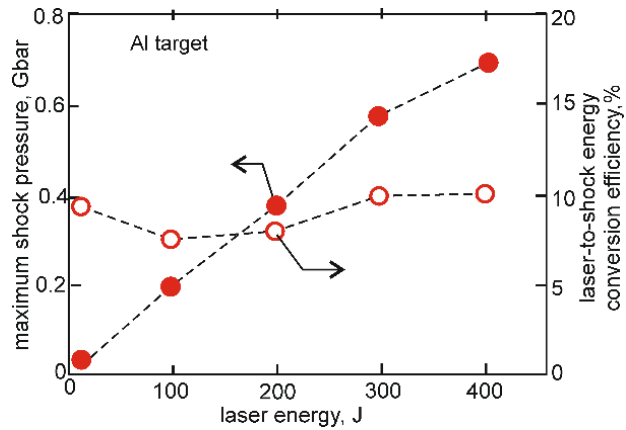

Fig. 12. The dependence of the maximum (in space and time) pressure of the shock and the laser-to-shock energy conversion efficiency in the Al target impacted by the gold projectile on the laser energy. The conversion efficiency was calculated for the moment t = $t_{pmax}$ when the shock pressure reaches the highest value.

Like in other methods of shock generation also in the method using the LICPA-based collider there are a number of factors that limit the achievable energy and pressure of the generated shock. One of the most important factors is the attainable velocity of the projectile accelerated in the collider. A natural barrier limiting this velocity is the sound speed of plasma in the cavity of the LICPA accelerator. Our numerical studies using the PALE code have shown that the temperature of plasma in the cavity can be by an order of magnitude higher than the temperature of plasma in the free space, which means that the speed of sound in the cavity plasma is ~ 3 times higher than in case of "free ablation". Thus, in the conditions examined the maximum velocity of the projectile in the LICPA scheme can be three times higher than in the AA scheme. The numerical value of the velocity limit depends on the accelerator and the driver's parameters as well as the initial parameters of the projectile and cannot be determined without knowing these parameters.

A significant role in limiting the attainable pressure of the shock produced by the projectile impact may play a spatial expansion of the projectile plasma during the projectile acceleration. This effect is crucial in the AA scheme and the CBA scheme where both the longitudinal and the transverse expansion of the plasma occurs and, as a consequence, the projectile density decreases quickly during acceleration. Due to the density decrease, despite the increase in velocity and energy of the projectile, the energy fluence and energy density of the projectile may not increase but decrease. Moreover, the time of interaction of the projectile with the impacted target may increase due to the increase of the thickness of the projectile during acceleration. As a result, the maximum pressure of the shock wave is achieved at a certain optimal distance/time of acceleration, and the value of this pressure is determined by the projectile's parameters attained at this distance (density, kinetic energy fluence, thickness). In the LICPA scheme only the longitudinal expansion of the projectile occurs and the projectile density decrease is much slower than for AA and CBA and the optimal time/distance of acceleration can be much longer. As a result, the attainable kinetic energy fluence of the projectile can be considerably higher than for AA and CBA (due to both the higher density and the higher velocity of the projectile) and the attainable pressure of the shock generated by the projectile impact can be considerably higher as well.

Other factors limiting parameters of the shock generated in the LICPA-based collider are hydrodynamic instabilities (Rayleigh-Taylor and possibly others) occurring during acceleration of the projectile. They can produce a large inhomogeneity in the density distribution of the projectile and, as a result, can be a source of large inhomogeneity in the shock generated by the projectile impact. However, one can expect that the destructive role of the hydro instabilities in the LICPA scheme can be manifested in stages of projectile acceleration latter than in the AA case because, due to much higher energetic efficiency of the acceleration in the LICPA scheme, we can use the accelerated target much thicker than in the AA scheme. The problem of hydro instabilities in the LICPA scheme has not been investigated by us in detail so far and will be the subject of our future research.

Although the factors limiting the attainable parameters of the projectile and the shock indicated above are generally the same for LICPA, AA and CBA, the upper limits for these parameters are different for each of the schemes and, at least for the laser energy range investigated so far (< 0.5 kJ), seem to be the highest for the LICPA scheme.

## 4. Conclusions

In conclusions, the generation of high-pressure shocks in solid targets by the laser-driven hyper-velocity impact in the recently proposed LICPA-based collider has been investigated using 2D hydrodynamic code. It has been shown that with the laser driver energy of only a few hundred joules, accessible for many existing one-beam nanosecond or sub-nanosecond laser facilities, sub-Gbar shock pressures can be produced in the collider with the laser-to-shock energy conversion efficiency by an order of magnitude higher than in other laser-based schemes used so far. In the laser energy range of 10 – 400 J, the peak shock pressure approximately linearly increases with the laser energy. The pressure value depends on both the projectile density and the impacted target density and it increases with an increase in these densities. The laser-to-shock energy conversion efficiency rather weakly depends on the laser energy; however, it pretty much depends on the target density – for a gold projectile driven by the laser pulse of energy of 200 J it reaches ~ 10 % for the Au target and ~ 20 % for the CH target. For all the target applied (Au, Cu, Al, CH) the pressure in the shock increases to its peak value very rapidly (within Δt < 0.3 ns for $E_L$ = 200 J) and no precursor disturbing the target material in front of the shock is observed.

The LICPA-based collider seems to be an efficient tool for the shock generation also in case of laser energies higher than those considered in this paper. In particular, it could be expected that with multi-kJ laser drivers the generation of strong shocks of multi-Gbar pressures will be feasible in the collider. To proof the usefulness of the collider in this laser energy range more studies are needed including detailed studies of hydrodynamic instabilities in the projectile acceleration process. On the other hand, using low-energy (~ 5 – 10 J) commercial nanosecond lasers for driving the collider, the shock pressures from several Mbar to several tens of Mbar could be quite easily produced. The last possibility would create the chance to conduct research in high energy-density science also in small, university-class laboratories.

**Acknowledgments**

This work was supported in part by the Ministry of Science and Higher Education, Poland under the grant no W39/7.PR/2015. Also, this work was supported in part by the Czech Ministry of Education, Youth and Sport , projects LD 14089 and RVO 68407700, as well as by the Eurofusion Consortium, project AWP15-ENR-01/CEA-02.